\documentclass[a4paper,11pt]{article}
\usepackage{pos}

\usepackage[
  per-mode=symbol-or-fraction,
]{siunitx}
\usepackage{braket}

\usepackage[section, below]{placeins}
\usepackage{subcaption}

\title{Inclusive semi-leptonic decays of charmed mesons with Möbius domain wall fermions}

\author*[a]{Ryan Kellermann}
\author[b, c]{Alessandro Barone}
\author[a]{Shoji Hashimoto}
\author[b, c, d]{Andreas Jüttner}
\author[a, e]{Takashi Kaneko}

\affiliation[a]{Theory Center, Institute of Particle and Nuclear Studies, High Energy Accelerator Research Organization (KEK), Tsukuba 305-0801, Japan and School of High Energy Accelerator Science, The Graduate University
for Advanced Studies (SOKENDAI), Tsukuba 305-0801, Japan}

\affiliation[b]{School of Physics and Astronomy, University of Southampton, Southampton SO17 1BJ, United Kingdom}

\affiliation[c]{STAG Research Center, University of Southampton, Southampton SO17 1BJ, UK}

\affiliation[d]{CERN, Theoretical Physics Department, Geneva, Switzerland}

\affiliation[e]{Kobayashi-Maskawa Institute for the Origin of Particles and the Universe, Nagoya University, Aichi 464–8602, Japan}

\emailAdd{kelry@post.kek.jp}

\abstract{We perform a non-perturbative lattice calculation of the decay rates for inclusive semi-leptonic decays of charmed mesons. In view of the long-standing tension in the determination of the CKM matrix elements $|V_{ub}|$ and $|V_{cb}|$ from exclusive and inclusive processes, recently, the use of lattice QCD has been extended towards the description of inclusive decays. Since the determination of hadronic input parameters from QCD based methods require independent tests, we focus on the charm sector, since it not only offers experimental data, but also well determined CKM parameters.
We carry out a pilot lattice simulation for the $D_s \rightarrow X_s \ell\nu$ and explore the improvement of existing techniques. Our simulation employs Möbius domain-wall charm and strange quarks whose masses are tuned to be approximately physical and we cover the whole kinematical region. We report on our progress in analyzing different sources of systematic effects, such as the extrapolation of the kernel function chosen for the Chebsyhev approximation as well as the influence on the analysis in the region close to the kinematical limit.}

\FullConference{%
  The 39th International Symposium on Lattice Field Theory (Lattice 2022), \\
  08.-13. August, 2022, \\
  Hörsaalzentrum Poppelsdorf, Bonn, Germany
}

\begin{document}
\maketitle

\section{Introduction}
In recent years, experiments have revealed a puzzling tension in B-decays, namely, in the determination of the CKM parameters $|V_{ub}|$ and $|V_{cb}|$ from exclusive and inclusive methods~\cite{Workman:2022ynf}. This discrepancy provides an opportunity for theorists to improve their 
understanding of these decays. Furthermore, the search for new physics requires precise theoretical predictions from the Standard Model. In view of these points, recently, ideas to extend the application of lattice QCD towards the description of inclusive decays have been proposed~\cite{Gambino:2020crt, Gambino:2022dvu, Hansen:2017mnd, Hansen:2019idp, Bulava:2021fre}. These approaches utilize either the Chebyshev approximation or the Backus-Gilbert approach to obtain the energy integral of the hadronic tensor, which defines the inclusive decay rates.

In this paper, we report on our progress in the application of this method towards a precise calculation of the inclusive semi-leptonic decay of the $D_s$-meson with a focus on presenting a method to control the systemtic error, which appears in the approximation of the kernel function in the energy integral.

First, we give a brief overview of the theoretical framework of our analysis and present the formulas used in the Chebyshev approximation. And secondly, we present our preliminary results of the analysis, as well as  a first, admittedly conservative, way to estimate the error in the limits that have to be taken to properly estimate the inclusive decay rate.

\section{Formulation of the Chebyshev approach}

We start with the definition of the total decay rate~\cite{Gambino:2020crt}
\begin{equation}
    \Gamma = \frac{G_F^2 |V_{cs}|^2}{24\pi^3} \int_{0}^{\boldsymbol{q^2_{\text{max}}}} d\boldsymbol{q}^2 \sqrt{\boldsymbol{q}^2} \bar{X}(\boldsymbol{q}^2) \, ,
    \label{equ:TotalRate}
\end{equation}
where we introduce the short-hand notation for the energy integral
\begin{equation}
    \bar{X} = \int_{\omega_{\text{min}}}^{\omega_{\text{max}}} d\omega \, K_{\mu\nu}(\boldsymbol{q},\omega) W^{\mu\nu} \, .
    \label{equ:OmegaInt}
\end{equation}
Here, $K_{\mu\nu}(\boldsymbol{q},\omega)$ is a kinematical factor given by the leptonic tensor and $W^{\mu\nu}$ is the hadronic tensor given by
\begin{equation}
    W^{\mu\nu}(p,q) = \sum_{X_s} (2\pi)^3 \delta^{(4)}(p-q-r) \frac{1}{2E_{D_s}}\braket{D_s(\boldsymbol{p})|\tilde{J}^{\mu\dagger}(-\boldsymbol{q})|X_s(\boldsymbol{r})}\braket{X_s(\boldsymbol{r})|\tilde{J}^{\nu}(\boldsymbol{q})|D_s(\boldsymbol{p})} \, ,
    \label{equ:HadronicTensor}
\end{equation}
where we sum over all possible final states $X_s$ to represent the inclusive decay and $\tilde{J}^{\nu}(\boldsymbol{q})$ is the Fourier transform of the inserted current, defined through $\tilde{J}^{\nu}(\boldsymbol{q}) = \sum_{\boldsymbol{x}} e^{-i\boldsymbol{q}\cdot\boldsymbol{x}} J^{\nu}(x)$.\\
The energy integral in \eqref{equ:OmegaInt} can be rewritten as
\begin{equation}
    \bar{X} = \braket{D_s(\boldsymbol{p})|\tilde{J}^{\mu\dagger}(-\boldsymbol{q})K_{\mu\nu}(\boldsymbol{q},\hat{H})\tilde{J}^{\nu}(\boldsymbol{q})|D_s(\boldsymbol{p})} \, ,
    \label{equ:AnalyticRes}
\end{equation}
where all intermediate states may contribute between the currents.
On the lattice, we calculate four-point correlation functions, which can be used to extract the matrix element
\begin{equation}
    C_{JJ} ^{\mu\nu}(\boldsymbol{q},t) = \frac{1}{V} \frac{1}{2m_{D_s}} \braket{D_s|\tilde{J}_\mu^\dagger(-\boldsymbol{q}) e^{-\hat{H}t} \tilde{J}_\nu(\boldsymbol{q})|D_s} \, .
    \label{equ:LatticeRes}
\end{equation}
We choose the rest frame of the initial $D_s$ meson, i.e. $\boldsymbol{p}=0$.\\

By comparing \eqref{equ:AnalyticRes} and \eqref{equ:LatticeRes}, we see that we can obtain $\bar{X}$ once an approximation of $K_{\mu\nu}(\boldsymbol{q},\hat{H})$ in terms of $e^{-\hat{H}}$ of the form
\begin{align*}
      K(\boldsymbol{q},\hat{H}) = k_{0} + k_{1} e^{-\hat{H}} + ... + k_{N} e^{-N\hat{H}} \, ,
\end{align*}
can be constructed as it allows to create an approximation of the energy integral \eqref{equ:AnalyticRes}
\begin{align*}
      \bar{X} \sim \, &k_{0} \underbrace{\braket{D_s|\tilde{J}_\mu^\dagger(-\boldsymbol{q})\tilde{J}_\nu(\boldsymbol{q})|D_s}}_{C_{\mu\nu}^{JJ}(0)} + k_{1} \underbrace{\braket{D_s|\tilde{J}_\mu^\dagger(-\boldsymbol{q})e^{-\hat{H}}\tilde{J}_\nu(\boldsymbol{q})|D_s}}_{C_{\mu\nu}^{JJ}(1)} + ... \\ &+ k_{N} \underbrace{\braket{D_s|\tilde{J}_\mu^\dagger(-\boldsymbol{q})e^{-\hat{H}N}\tilde{J}_\nu(\boldsymbol{q})|D_s}}_{C_{\mu\nu}^{JJ}(N)} \, ,
\end{align*}
where the matrix elements on the right hand side are determined by the lattice data.
In our case, we employ the shifted Chebyshev polynomials $T_j^*(e^{-\omega})$ to create an approximation of $K(\hat{H})$ in the integration range $[\omega_0, \infty]$ with $0 \leq \omega_0 < \omega_{\text{min}}$. 

\begin{figure}
    \begin{subfigure}{0.49\textwidth}
        \centering
        \includegraphics[width=\textwidth]{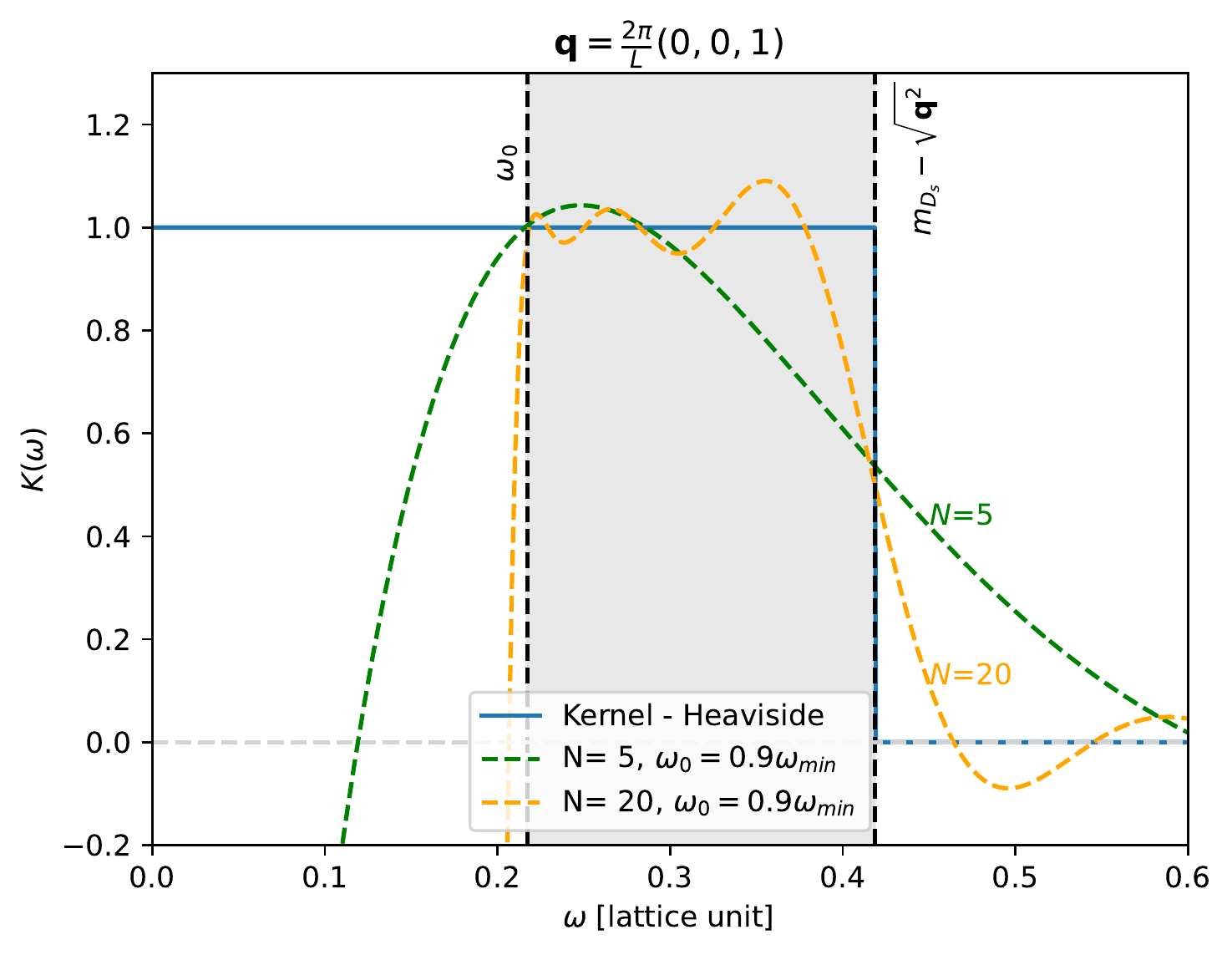}
        \caption{Heaviside function.}
        \label{fig:ApproximationHeavi}
    \end{subfigure}
    \begin{subfigure}{0.49\textwidth}
        \centering
        \includegraphics[width=\textwidth]{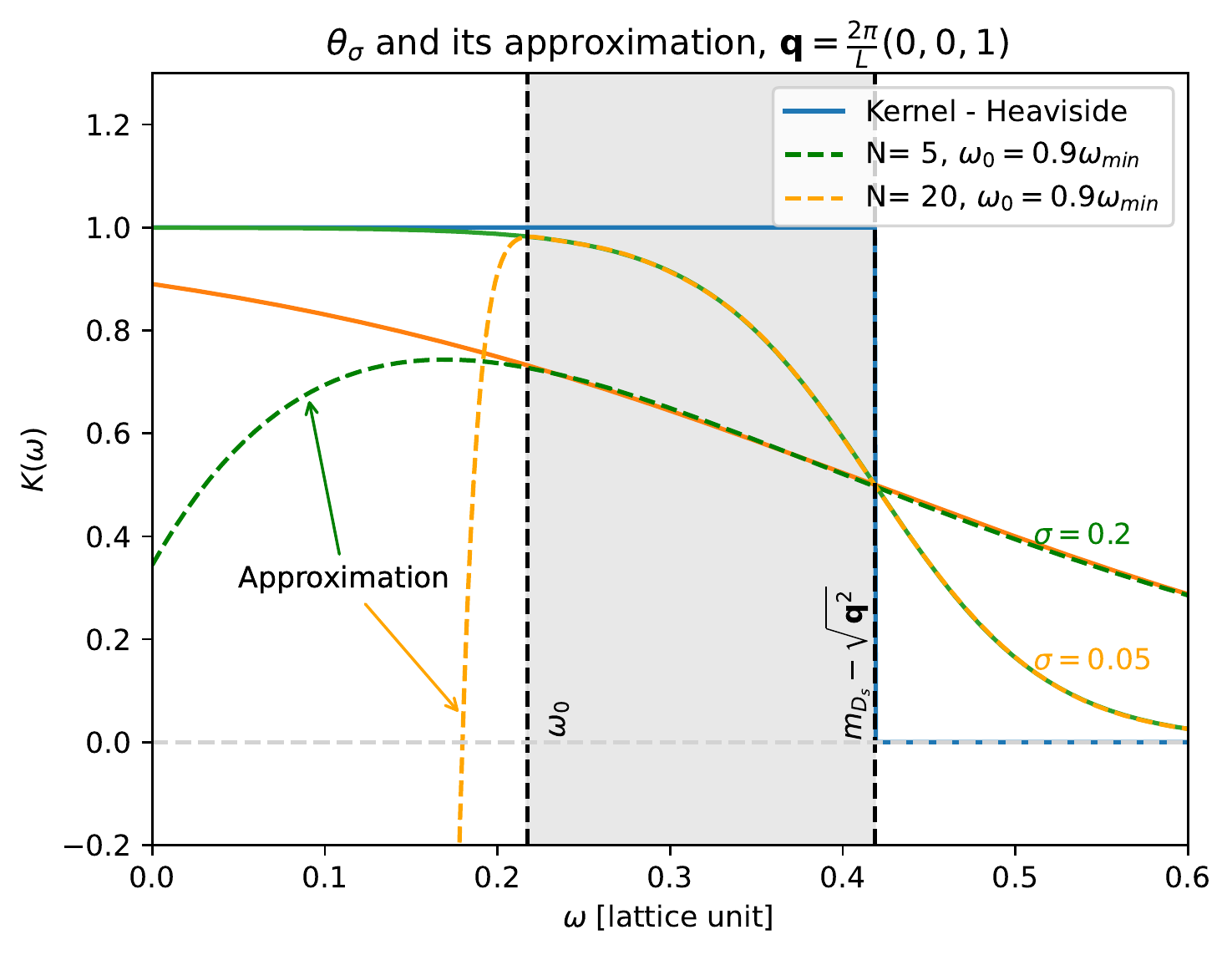}
        \caption{Sigmoid function.}
        \label{fig:ApproximationSmeared}
    \end{subfigure}
    \caption{Chebyshev approximation of the kernel function, depending on the choice of the kernel function. The blue solid line represents the Heaviside function on both sides. On the left hand side, the dashed lines represent a direct approximation of the Heaviside function. On the right hand side, the colored solid lines show the Sigmoid function defined in Eq. \eqref{equ:Sigmoid}, and the dashed lines show their approximation. We use two choices of $N$ in both plots and the smearing used on the right hand side plot is related to the number of polynomials via $\sigma = 1/N$.}
    \label{fig:ApproximationKernel}
\end{figure}

In the following, we show the behavior of the Chebyshev approximation depending on the choice of the kernel function. This is shown in Figure \ref{fig:ApproximationKernel}. First, we consider the kernel function $K(\boldsymbol{q},\omega)$ of a simple Heaviside function
\begin{equation}
    K(\omega) = \theta(m_{D_s} - \sqrt{\boldsymbol{q}^2} - \omega) \, ,
\end{equation}
to implement the upper limit of the $\omega$ integral. The approximation results are shown in Figure \ref{fig:ApproximationHeavi} and we see that by simply increasing the number of polynomials in the Chebyshev approximation from 5 to 20 increases the oscillations of the approximation. In order to stabilize the approximation, we smear the kernel function by introducing a smearing parameter $\sigma$, i.e we employ a Sigmoid function of the form
\begin{equation}
    \theta_\sigma(m_{D_s} - \sqrt{\boldsymbol{q}^2} - \omega) \equiv \frac{1}{1+e^{-\frac{m_{D_s} - \sqrt{\boldsymbol{q}^2} - \omega}{\sigma}}} \, ,
    \label{equ:Sigmoid}
\end{equation}
and the approximation result is shown in Figure \ref{fig:ApproximationSmeared}. While this approach allows for a "smoother" approximation, it now requires us to take the limit of $\sigma \rightarrow 0$ in addition to $N \rightarrow \infty$ to obtain a proper estimate. This source of systematical error is the focus of this work.

To finalize this section, let us write down how we construct our approximation. The $\omega$-integral can be approximated by
\begin{align}
      \frac{\braket{\psi_\mu|K(\hat{H})|\psi_\nu}}{\braket{\psi_\mu|\psi_\nu}} = \frac{c^*_0}{2} + \sum_{j=1}^{N} c_j^* \underbrace{\frac{\braket{\psi_\mu|T^*_j(e^{-\hat{H}})|\psi_\nu}}{\braket{\psi_\mu|\psi_\nu}}}_{C(t+2t_0)/C(2t_0)} \, ,
      \label{equ:ChebyshevApprox}
\end{align}
where we define the state $\ket{\psi_\nu} \equiv e^{-\hat{H}t_0}\tilde{J}_\nu(\boldsymbol{q})\ket{D_s(\boldsymbol{0})}$ on which the kernel operator is evaluated. The $T^*_j(x)$ are the shifted Chebyshev polynomials, which can be obtained from the standard Chebyshev polynoials as $T^*_j(x) \equiv T_j(2x-1)$. The first few terms are given by $T^*_0(x) = 1$, $T^*_1(x) = 2x-1$, $T^*_2(x) = 8x^2 -8x +1$, and higher orders are obtained recursively thorugh $T^*_j(x) = (2x-1)T^*_{j-1}(x) - T^*_{j-2}(x)$. The coefficients $c_j^*$ depend on the choice of the lower limit of the $\omega$-integral $\omega_0$ and in the case of $\omega_0 = 0$ are simply given by
\begin{align}
       c^*_j = \frac{2}{\pi} \int_{0}^{\pi} d\theta K\left(-\ln\frac{1+\cos\theta}{2}\right) \cos(j\theta) \, .
\end{align}
An important property of the Chebyshev polynomials is that the Chebyshev matrix elements are confined between $[-1,1]$, i.e.
\begin{align}
       \left|\frac{\braket{\psi_\mu|T^*_j(e^{-\hat{H}})|\psi_\nu}}{\braket{\psi_\mu|\psi_\nu}}\right| \leq 1 \, .
       \label{equ:ChebyshevProperty}
\end{align}
This property can be used in two ways. Firstly, we can use it to suppress the statistical noise for high orders of $j$ in the Chebyshev approximation where we expect huge cancellation among different orders of $x$. And secondly, it allows us to estimate the upper limit of the error, since all Chebyshev matrix elements are bounded by $\pm1$ for any $j$.

\section{Numerical results}
This computation is performed on the lattice data generated with $2+1$-flavor Möbius domain wall fermions (ensemble "$M{\text -}ud3{\text -}sa$" from \cite{Colquhoun:2022atw}, which has $1/a = \SI{3.610(9)}{\giga\electronvolt}$). The charm and strange quarks are simulated at near physical values, while the up and down quark are simulated at a pion mass of $m_\pi \simeq \SI{300}{\mega\electronvolt}$.
The lattice volume is $48^3 \times 96$ and the forward-scattering matrix elements are calculated for spatial momenta $\boldsymbol{q}$ of $(0,0,0)$, $(0,0,1)$, $(0,1,1)$ and $(1,1,1)$ in units of $2\pi/L$. All the data have been generated with Grid \cite{Grid:Boyle} and Hadrons \cite{antonin_portelli_2022_6382460} software packages. Part of the fits in the analysis has been performed using lsqfit \cite{lepage:lsqfit}. 

The number of configurations averaged are 50 and the measurement is duplicated with 8 different source time slices.
For each fixed spatial momentum $\boldsymbol{q}$ we calculate the four-point correlation function to extract $C_{JJ}^{\mu\nu}(t,\boldsymbol{q})$ (further details on the lattice calculation can be found in \cite{Hashimoto:2017wqo}) and determine the shifted Chebyshev matrix elements from $C_{JJ}^{\mu\nu}(t+2t_0,\boldsymbol{q})/C_{JJ}^{\mu\nu}(2t_0,\boldsymbol{q})$ as shown in \eqref{equ:ChebyshevApprox} by performing a constrained fit imposing the condition \eqref{equ:ChebyshevProperty}. The $\omega$-integral is then obtained by using the representation \eqref{equ:ChebyshevApprox}.

\begin{figure}
    \centering
    \includegraphics[width=0.7\textwidth]{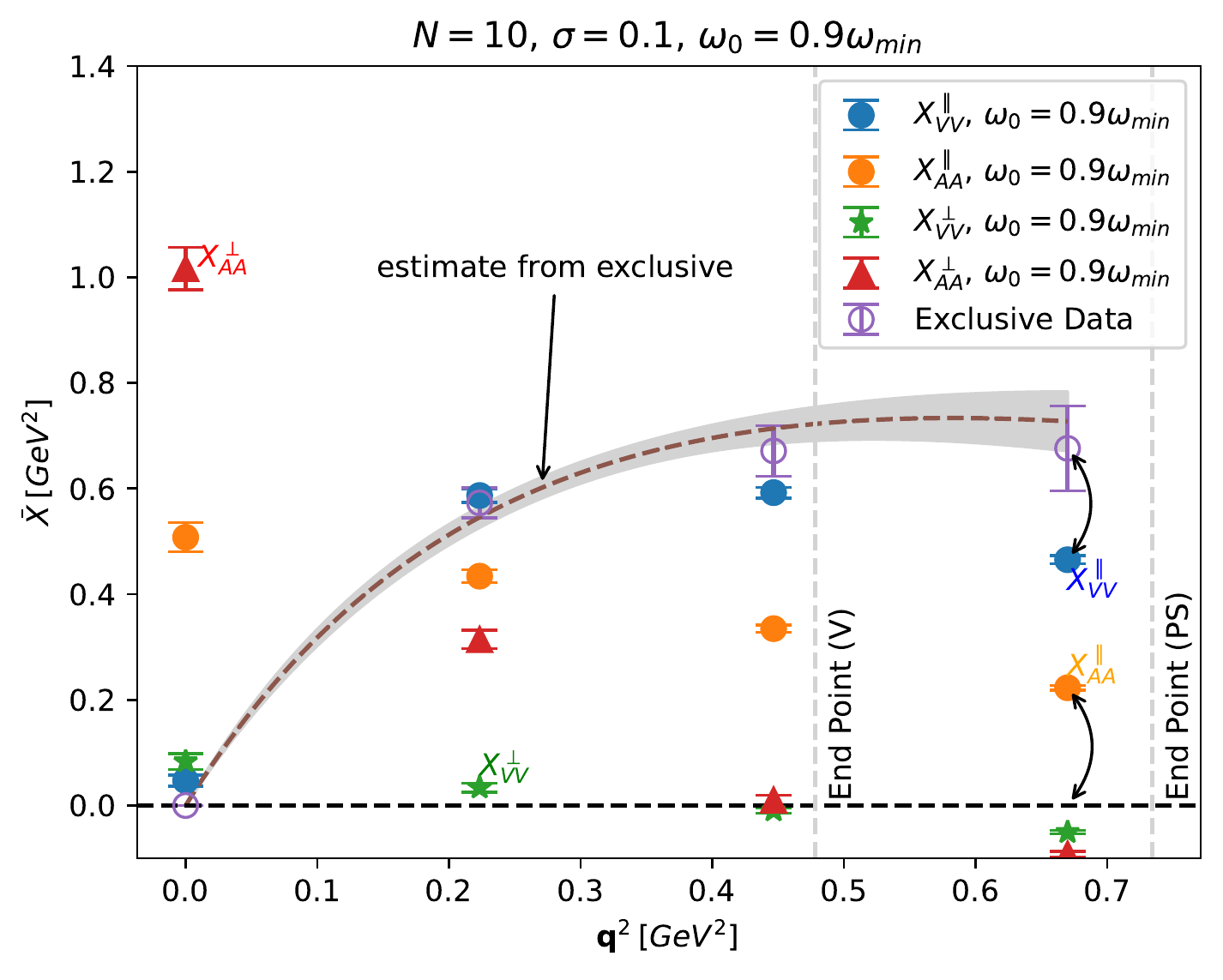}
    \caption{$\bar{X}$ contributions for different kinematical channels as a function of $\boldsymbol{q}$. The vertical lines show the value $\boldsymbol{q}^2_{\text{max}}$ for the vector (V) and pseudoscalar (PS) meson, respectively. The approximation is obtained for $N=10$ Chebyshev polynomials and the smearing of the kernel function is given by $\sigma = 1/N=0.1$. With the available lattice data $N=10$ is the upper limit for the Chbeyshev approximation, since the statistical noise of the data becomes too strong for higher orders.}
    \label{fig:XBar}
\end{figure}

In Figure \ref{fig:XBar} we show the preliminary results for the energy integral $\bar{X}$ defined in \eqref{equ:AnalyticRes}, where we decompose $\bar{X}$ into different contributions, i.e. whether we have vector (VV) or axial-vector (AA) current insertions, as well as the polarization of the inserted currents, i.e. parallel ($\parallel$) and perpendicular ($\perp$) to the momentum $\boldsymbol{q}$. Our results are shown for a choice of $N=10$ and the smearing of the kernel function \eqref{equ:Sigmoid} is defined through $\sigma = 1/N = 0.1$. With the available data, $N=10$ is the highest order that we can achieve with the Chebyshev approximation, since the statistical noise of the lattice data becomes too large for orders of $N > 10$. In Figure \ref{fig:XBar}, we also include a contribution to $\bar{X}_{VV}^{\parallel}$ from the exclusive semi-leptonic $D \rightarrow K$ decay, allowing us to surmise that our results are in the right ballpark.
\FloatBarrier
We comment on the region close to the end of the phase space, i.e. the point of $\boldsymbol{q}=(1,1,1)$ corresponding to $\boldsymbol{q}^2 \approx \SI{0.66}{\giga\electronvolt^2}$ shown in Figure \ref{fig:XBar}, for $X^{\parallel}_{VV}$ and $X^{\parallel}_{AA}$.
In this region, a dominant contribution from the ground state is expected for $X^{\parallel}_{VV}$, since the excited state energy exceeds the $D_s$ meson mass, while the expected contribution to $X^{\parallel}_{AA}$ should already be zero, because the lowest energy state ($s\bar{s}$-vector) has an energy lager than $m_{D_s}$. Figure \ref{fig:XBar} shows large discrepancies between these expected values (dashed lines) and our approximation (data points).

\section{Study of systematic errors}

\subsection{Above the kinematical end-point: $X_{AA}^{\parallel}$}

First, we consider the case of $X_{AA}^{\parallel}$. In this case we expect contributions from the vector meson in the final state. At $\boldsymbol{q} = (1,1,1)$, the energy of the lowest state is already above the threshold, so that the expected result is zero. For a finite order of the polynomials $N=10$ and a non-zero smearing width $\sigma=0.1$, Figure \ref{fig:XBar} shows that this is not the case.

\begin{figure}
    \centering
    \includegraphics[width=0.7\textwidth]{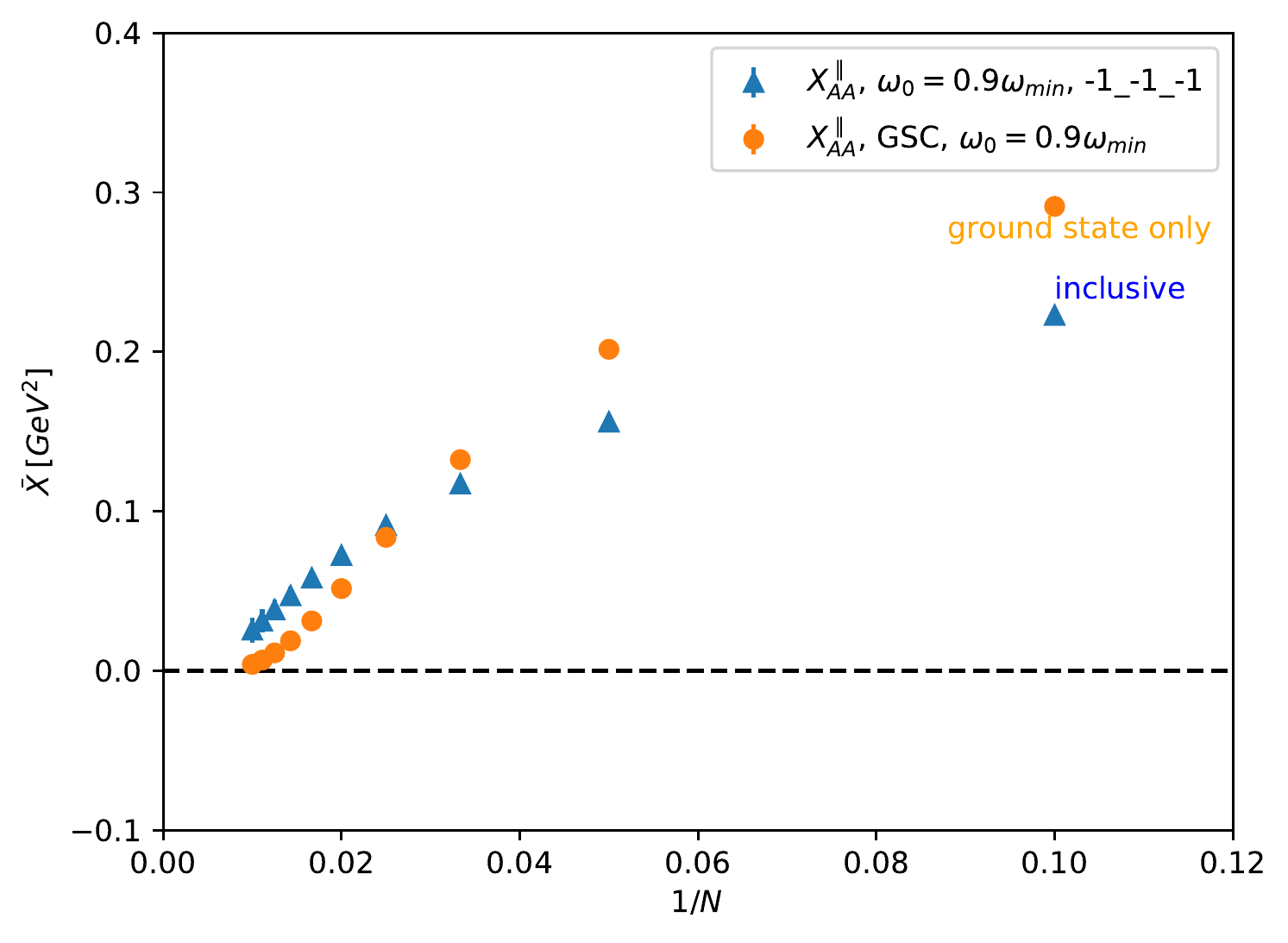}
    \caption{Development of approximation results for $X_{AA}^{\parallel}$ at $\boldsymbol{q}=(1,1,1)$ depending on the number of polynomials $N$ used in the Chebyshev approximation. We set the smearing $\sigma$ of the kernel function to be $\sigma = 1/N$. The blue triangles show the approximation results using the available lattice data, while the orange circles are obtained by only considering the ground state contribution obtained from fitting the lattice data.}
    \label{fig:XAAParallel}
\end{figure}

To take both, $\sigma \rightarrow 0$ and $N\rightarrow \infty$, limits simultaneously, we set $\sigma = 1/N$ and consider the evolution of our approximation as a function of $1/N$. The result is shown in Figure \ref{fig:XAAParallel}, where $N$ is taken to be between 10 and 100. To access the Chebyshev matrix elements of orders higher than $N =10$, we use the property \eqref{equ:ChebyshevProperty} of the Chebyshev matrix elements, i.e. the fact that the Chebyshev matrix elements are bound by $0\pm1$. It allows us to simply add up the absolute values of the Chebyshev coefficients for $j > 10$ in \eqref{equ:ChebyshevApprox} to obtain a mathematical upper limit of the error for any given order of $N$. 
Additionally, we include the result of our approximation in that we only consider the ground state contribution. This estimate is obtained by fitting the lattice data and the extracted ground state energy is used to calculate the Chebyshev matrix elements up to arbitrary order. We can see that even for the limited number of polynomials, say $N=100$, the approximation approaches zero sufficiently fast.
\FloatBarrier
\subsection{Above the kinematical end-point: $X_{VV}^{\parallel}$}

\begin{figure}
    \centering
    \includegraphics[width=0.7\textwidth]{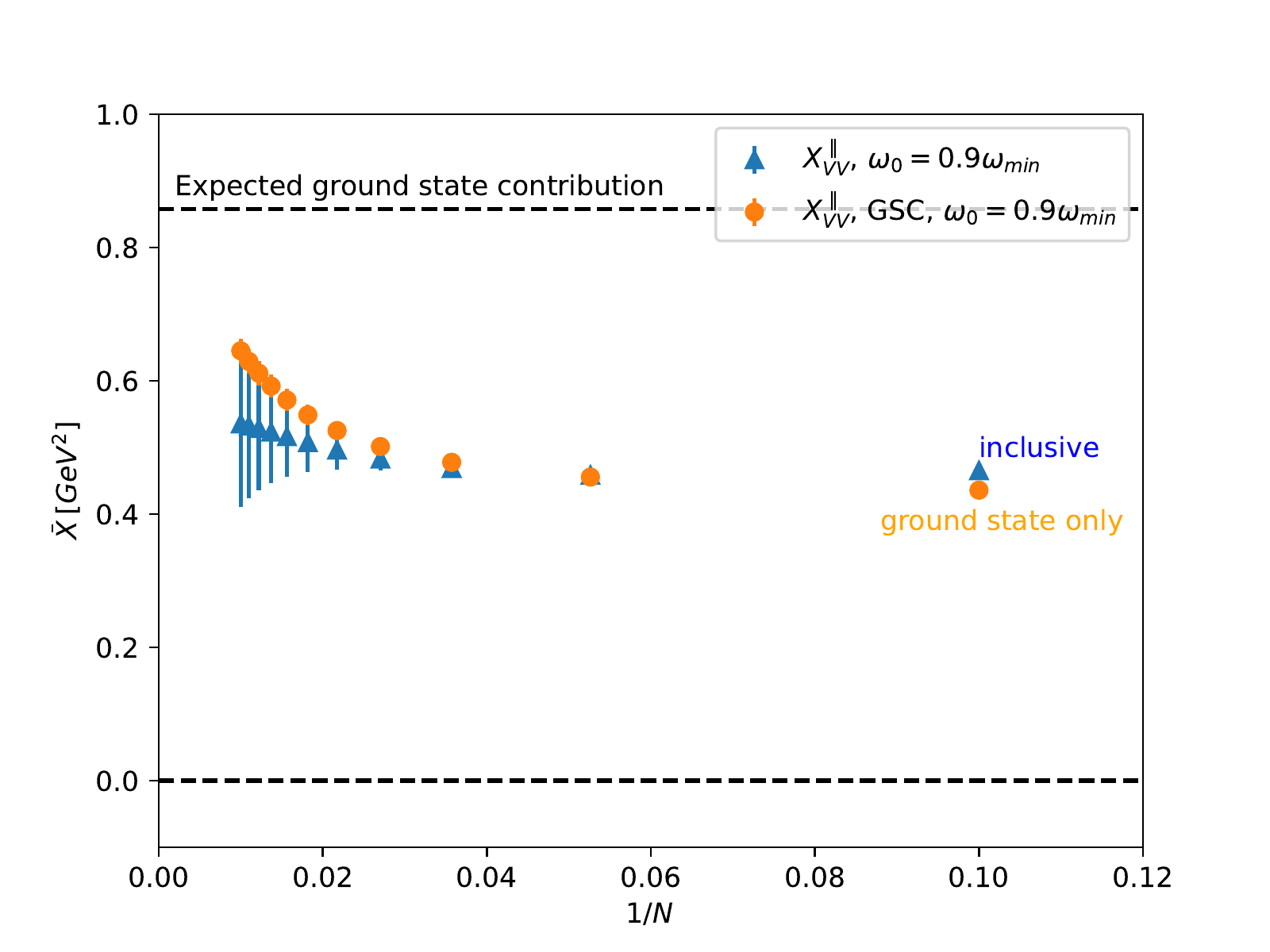}
    \caption{Development of approximation results for $X_{VV}^{\parallel}$ at $\boldsymbol{q}=(1,1,1)$ depending on the number of polynomials $N$ used in the Chebyshev approximation. We set the smearing $\sigma$ of the kernel function to be $\sigma = 1/N$.The blue triangles show the approximation results using the available lattice data, while the orange circles are obtained by only considering the ground state contribution obtained from fitting the lattice data.}
    \label{fig:XVVParallel}
\end{figure}

\begin{figure}
    \centering
    \includegraphics[width=0.7\textwidth]{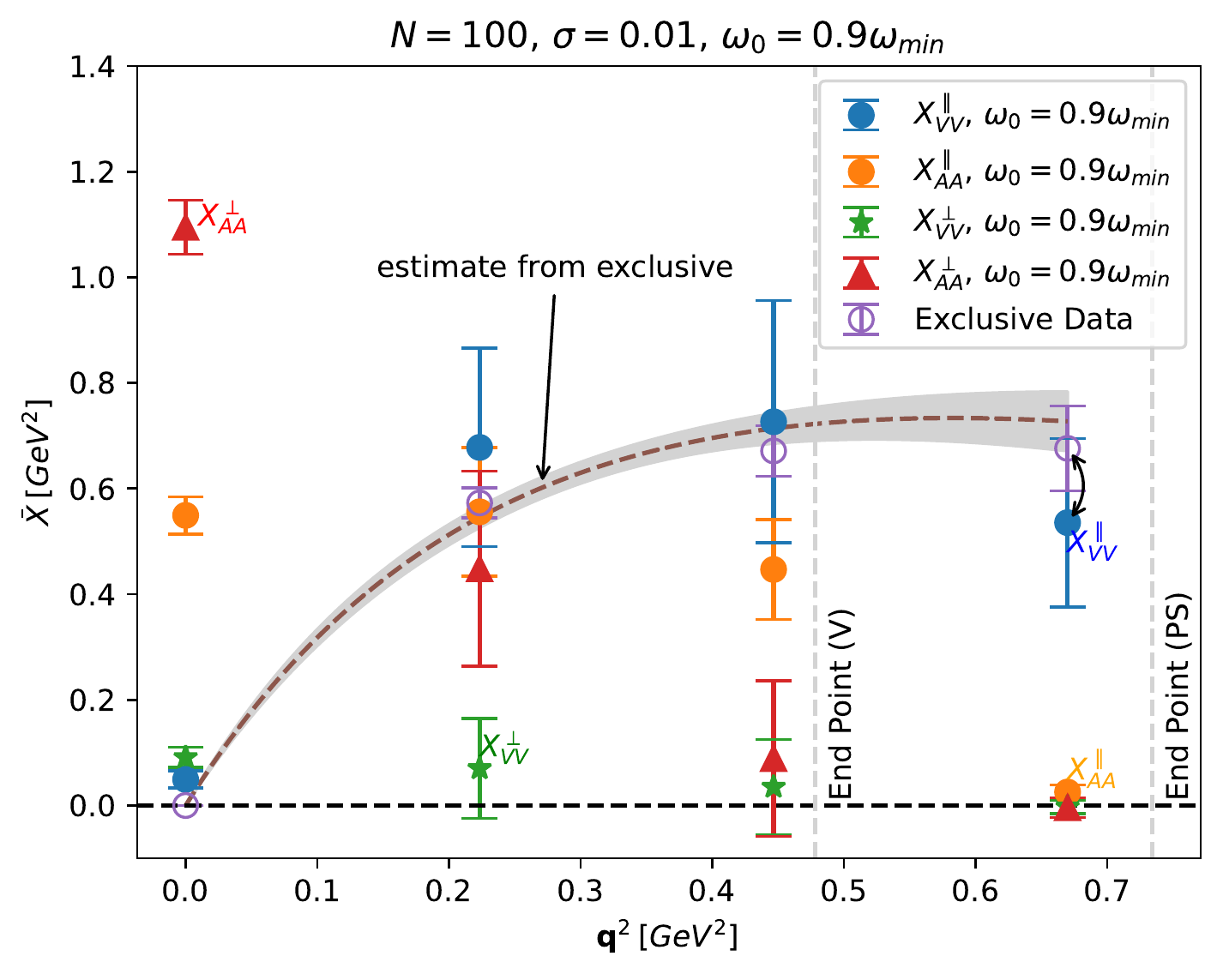}
    \caption{Chebyshev approximation for $\bar{X}$ given in Eq. \eqref{equ:OmegaInt} decomposed into different kinematical channels for $N=100$ Chebyshev polynomials. The smearing of the kernel function is given by $\sigma = 1/N = 0.01$. The error bars show the mathematical upper limit obtained by employing the properties of the Chebyshev matrix elements.}
    \label{fig:XBarComparison}
\end{figure}

For $X_{VV}^{\parallel}$ we expect contributions from both the pseudoscalar and vector mesons. We have to take both of these contributions into consideration to obtain an estimate of the ground-state-only contribution, which, together with the results obtained from using the inclusive data, are shown in Figure \ref{fig:XVVParallel}. Here, it is important to note how the error on the inclusive data is estimated. Taking into account the analytical form of the approximation given in Eq. \eqref{equ:ChebyshevApprox} and the fact that for higher orders of $N$ the Chebyshev matrix elements are basically given by $0\pm1$, we can construct an error estimate by simply adding up the absolute values of the coefficients $c_j^*$ appearing in the approximation. These error estimates are shown in Figure \ref{fig:XVVParallel}.
The figure shows that our error estimate is able to cover the expected ground state contribution. Furthermore, the behavior of the ground state contribution, i.e. the steady increase of the approximation value, shown in the Figure \ref{fig:XVVParallel} is expected. For $\bar{X}_{VV}^{\parallel}$ at $\boldsymbol{q} = (1,1,1)$ the range of the energy integral is quite narrow and this range is dominated by the ground state. So that depending of the choice of the Chebyshev polynomials $N$, and consequently the smearing $\sigma$, our approximation monotonously increases towards the true value.
\FloatBarrier
Finally, let us close this section by showing how the results shown in Figure \ref{fig:XBar} change if we increase $N$ to 100 and apply the error estimation method discussed above. The results are shown in Figure \ref{fig:XBarComparison}. The Figure shows that even if the number of polynomials is increased the central values of the approximation remains stable. This should also be the case if we take the $N \rightarrow \infty$ limit. At the same time we see that the error bars also start increasing significantly. We note that the error bars shown in this plot are most likely overestimated since we are assuming the mathematical upper limit. The actual error is expected to be smaller, but a proper estimate requires knowledge on the spectrum. For instance, with a flat spectrum, the errors should cancel around the threshold, while real problems might occur if the spectrum is rapidly changing, although this is only expected near the ground state.
\FloatBarrier
\section{Conclusion}

We reported on our progress towards a lattice computation of the inclusive $D \rightarrow X\ell\nu$ decay with fully controlled statistical effects. We focused on the systematical error arising due to the approximation of the kernel function and presented a conservative error estimate employing the mathematical properties of the Chebyshev approximation. With this estimate we are able to cover the expected ground state contribution for the region close to the kinematical limit where a ground state dominance is expected. To obtain more realistic error bars further study is required. Once a proper error estimate is available we will obtain estimates for the total decay rate and compare our results with experimental data~\cite{BESIII:2021duu}.
Furthermore, we are going to extend our analysis by including two more ensembles, as well as considering different inclusive channels.

\section*{Acknowledgments}

The numerical calculations of the JLQCD collaboration were performed on SX-Aurora TSUBASA at the High Energy Accelerator Research Organization (KEK) under its Particle, Nuclear and Astrophysics Simulation Program, as well as on Fugaku through the HPCI System Research Project (Project ID: hp220056).

The works of S.H. and T.K. are supported in part by JSPS KAKENHI Grant Numbers 22H00138 and 21H01085, respectively, and by the Post-K and Fugaku supercomputer project through the Joint Institute for Computational Fundamental Science (JICFuS).

\end{document}